\documentclass[a4paper]{jpconf}
\usepackage[english]{babel}
\usepackage[export]{adjustbox}
\usepackage{tikz,lipsum,enumitem,graphicx,array,hyperref,bm,csquotes,biblatex}
\usepackage{etoolbox}
\usepackage{amsmath}
\usepackage{xcolor}
\usepackage{color}
\usepackage{soul} 

\usetikzlibrary{chains,fit,arrows.meta,shapes.multipart,pgfplots.groupplots}
\usetikzlibrary{positioning,shapes,topaths,calc,backgrounds,matrix,shadows,automata,patterns,decorations.pathmorphing,decorations.pathreplacing,decorations.text,trees}
\pgfdeclarelayer{background}
\pgfsetlayers{background,main}
\usepackage{listing,lstautogobble}
\definecolor{lstkeyword}{HTML}{01497c}
\definecolor{lststring}{HTML}{a4133c}
\definecolor{lstcomment}{HTML}{718355}
\definecolor{lsthl}{HTML}{fcca46}
\lstdefinestyle{c++}{
    language=c++,
    escapechar=`,
    tabsize=3,
    backgroundcolor=\color{black!7},
    basicstyle={\ttfamily\small},
    keywordstyle={\color{lstkeyword}\bfseries},
    commentstyle={\color{lstcomment}},
    stringstyle={\color{lststring}},
}

\usepackage{pgfplots}
\pgfplotsset{
    compat=1.18,
    error bars/error bar style={line width=1pt,black},
    error bars/error mark options={line width=1pt,black,mark size=2pt,rotate=90},
    every mark/.append style={mark size=3pt},
}
\definecolor{palette0_0}{HTML}{61a5c2}
\definecolor{palette0_1}{HTML}{2a6f97}
\definecolor{palette0_2}{HTML}{013a63}
\definecolor{palette0_3}{HTML}{ffba08}
\definecolor{palette0_4}{HTML}{e85d04}
\definecolor{palette0_5}{HTML}{6a040f}
\pgfplotscreateplotcyclelist{multiline1}{
    {palette0_0,thick,mark=x},
    {palette0_1,thick,mark=square*},
    {palette0_2,thick,mark=triangle*},
    {palette0_4,thick,mark=*},
}
\definecolor{palette1_0}{HTML}{4d908e}
\definecolor{palette1_1}{HTML}{003554}
\definecolor{palette1_2}{HTML}{006494}
\definecolor{palette1_3}{HTML}{0582ca}
\definecolor{palette1_4}{HTML}{00a6fb}
\pgfplotscreateplotcyclelist{perfplot}{
    {draw=palette1_0,fill=palette1_0},
    {draw=palette1_1,fill=palette1_1},
    {draw=palette1_2,fill=palette1_2},
    {draw=palette1_3,fill=palette1_3},
    {draw=palette1_4,fill=palette1_4},
}

\begin{document}
\title{RNTuple: Towards First-Class Support for HPC data centers}
\author{Giovanna Lazzari Miotto$^1$, %
Javier Lopez-Gomez$^2$}
\address{$^1$ Universidade Federal do Rio Grande do Sul (BR)\\ %
$^2$ EP-SFT, CERN, Geneva, Switzerland}
\ead{$^1$ glmiotto@inf.ufrgs.br, $^2$ javier.lopez.gomez@cern.ch}

\begin{abstract}
Compared to LHC Run 1 and Run 2, future HEP experiments, e.g., at the HL-LHC, will increase the volume of generated data by an order of magnitude. In order to sustain the expected analysis throughput, ROOT's RNTuple I/O subsystem has been engineered to overcome the bottlenecks of the TTree I/O subsystem, focusing also on a compact data format, asynchronous and parallel requests, and a layered architecture that allows supporting distributed filesystem-less storage systems, e.g. HPC-oriented object stores.
In a previous publication, we introduced and evaluated the RNTuple's native backend for Intel DAOS. Since its first prototype, we carried out a number of improvements both on RNTuple and its DAOS backend aiming to saturate the physical link, such as support for vector writes and an improved RNTuple-to-DAOS mapping, only to name a few. In parallel, the latest developments allow for better integration between RNTuple and ROOT's storage-agnostic, declarative interface to write HEP analyses, RDataFrame.
In this work, we contribute with the following: \textit{(i)} a redesign of the RNTuple DAOS backend, including a mechanism for efficient population of the object store based on existing data; and \textit{(ii)} an experimental evaluation on a single-node platform, showing a significant increase in the analysis throughput for typical HEP workflows.
\end{abstract}

\section{Introduction}
In the last few decades, the ROOT TTree~\cite{brun1997root} I/O has been used to efficiently store more than one exabyte of high-energy physics data and became the HEP de-facto standard format. Its on-disk columnar layout allows for efficient reading of a subset of the columns, a common case in HEP analyses. However, future experiments, e.g., in the HL-LHC, are expected to increase the size of datasets by one order of magnitude, which makes high-bandwidth low-latency NVMe/SCM memory and distributed object stores especially relevant. TTree cannot fully exploit the capabilities of the aforementioned storage systems.
RNTuple~\cite{EPJrntuple2020} is the new columnar I/O subsystem for ROOT that addresses TTree's shortcomings. In particular, it targets high performance on low-latency NVMe devices, asynchronous and concurrent bulk I/O, native support for object stores, e.g., DAOS, and robust, user-friendly interfaces.

Our contribution in this paper can be summarized as follows:
\begin{itemize}
    \item We propose a matured, RNTuple--DAOS backend that delivers significantly improved read/write throughput. We recognize the importance of write throughput given that data must be imported into the object store in order to be processed in HPC facilities.
    \item We provide an experimental evaluation that demonstrates the performance improvements of the proposed techniques in a realistic, end-to-end analysis.
\end{itemize}

\section{Background}
In this section, we introduce a number of key concepts that relate to both ROOT's RNTuple I/O subsystem and Intel DAOS as the used object store.

\subsection{ROOT's RNTuple I/O subsystem}
\begin{figure}[!hb]
    \centering\vspace{-1em}
    \resizebox{.88\textwidth}{!}{\input{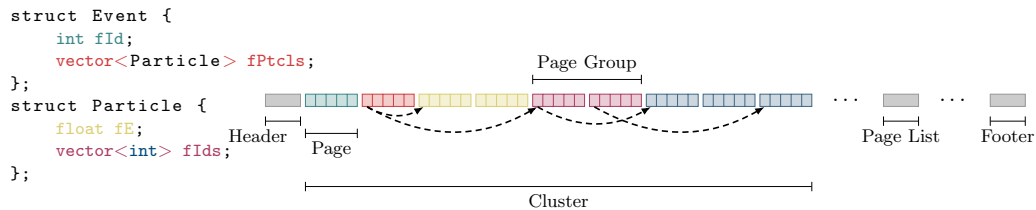}}\vspace{-1em}
    \caption{RNTuple's on-disk layout. Pages store data for the struct member of matching color.}
    \label{fig:ntuple_ondisklayout}
\end{figure}

In RNTuple, data is stored column-wise on disk, similarly to TTree and Apache Parquet~\cite{Vohra2016}. An RNTuple file consists of the actual stored data and metadata structures that describe how the data is to be interpreted. User data is organized into pages and clusters. A page contains an array of contiguous values of a fundamental data type (e.g., \texttt{Int32}) for a given column, whereas clusters hold all the pages containing data for a range of rows in the dataset. Pages for the same column and cluster denote a page group. On-disk pages are usually compressed. Columns of a complex C++ type, e.g., \verb|std::vector<float>|, are broken up into columns of fundamental data types. Metadata encompasses the header, the page lists, and the footer which contain, respectively, the schema description, the location of pages and clusters, and summary information and other metadata. For an illustrative example of the on-disk layout, see Figure~\ref{fig:ntuple_ondisklayout}.

RNTuple's design decouples data representation from raw storage of byte ranges, therefore making it possible to implement backends for different storage systems, such as POSIX files or object stores.

\vspace{-1em}
\subsection{DAOS}
DAOS~\cite{10.1007/978-3-030-48842-0_3} is part of the Intel exascale storage stack, and provides a fault-tolerant distributed object store targeting high bandwidth, low latency, and high IOPS provided by NVMe and SCM devices. DAOS completely bypasses the operating system virtual filesystem and block I/O layers. On the one hand, this forgoes kernel I/O coalescing and buffering, which is mostly relevant for spinning disks. On the other hand, the POSIX strong consistency model is known to limit the scalability of parallel filesystems. A DAOS system is comprised of multiple servers running a Linux daemon that exports local NVMe/SCM storage. RDMA is used where available, e.g., over InfiniBand, to copy data from servers to clients.

The storage space can be partitioned into pools, which is further divided into container namespaces, which hold objects. An object is a key--value store that is identified by a 128-bit object ID (\textit{oid}). DAOS splits its keys in two parts: the distribution key (\textit{dkey}) and the attribute key (\textit{akey}); in particular, the former impacts data locality by determining the physical device in which the data is stored.

\section{Efficient storage of HENP data in DAOS}\label{sec:proposal}
In previous publications~\cite{L_pez_G_mez_2021}, we introduced a prototype of a DAOS backend for RNTuple. However, this initial version did not fully exploit the data link. In the following, we describe a number of improvements that we carried out in order to improve not only the read throughput, but also shorten the time required to import a dataset into the object store.

Orthogonal to the changes described below, and given the relatively high cost of creating an operation queue, the backend was reworked to initialize a single, persistent queue per accessed ntuple. Beforehand, queues were created and destroyed for each vector read/write.

\subsection{Co-locality-based mapping function}\label{sec:proposal:mapping}
Guided by typical HENP analysis patterns, RNTuple data is designed to be accessed on a columnar, page group basis. However, the migration to an object store paradigm can disband related pages. Although previous mappings between RNTuple and DAOS did not exploit it (see Eqn.~\ref{eqn:oid_per_page}), the locality-affecting $dkey$ imparts a way to retain columnar semantics by crafting a mapping function that keeps page groups together (see Eqn.~\ref{eqn:locality-driven}). This unveils an opportunity to read data efficiently in parallel, provided that neighboring pages' requests are coalesced in advance.
In the equations below, $Cluster_i$ and $Column_j$ denote, respectively, the cluster and column identifiers, whereas $Page_k$ is a strictly increasing integer that uniquely identifies a page.
\begin{align}
    \boldsymbol{\phi}\colon \langle Cluster_i, Column_j, Page_k \rangle & \rightarrow\mathbf \langle oid, dkey, akey \rangle \nonumber\\
    \label{eqn:oid_per_page}
    \boldsymbol{\phi}_{object-per-page}(Cluster_i, Column_j, Page_k)    & \mapsto \langle Page_k, \alpha_{dkey}, \alpha_{akey}\rangle  \\
    \label{eqn:locality-driven}
    \boldsymbol{\phi}_{locality-driven}(Cluster_i, Column_j, Page_k) & \mapsto \langle Cluster_i, Column_j, Page_k\rangle
\end{align}

\subsection{Request coalescing}
In order to benefit from the parallel reads made possible by the locality-driven mapping in Section \ref{sec:proposal:mapping}, requests for coinciding pages in DAOS servers must be made in the same call. The DAOS's client API allows calls with an arbitrarily-long batch of requests for elements sharing the $oid$ and $dkey$. By deferring I/O calls until a cluster's requests have been coalesced by $\langle oid, dkey \rangle$, we realize the potential for much faster throughput and minimize operation queue overhead.

%

\subsection{Vector writes}
Though applicable to both reading and writing, the aforementioned improvements could not be fully leveraged by the I/O subsystem without support for vector writes, i.e., performing multiple update operations at once. Such batched transfers can improve throughput and amortize the overhead costs associated with the network fabric, and thus shorten the data import stage. 

RNTuple was already able to buffer pages and defer writing until the entire cluster is committed, but pages were still individually issued to the underlying backend. Now, the interface has been extended to trigger vector writes in storage backends that may benefit from it.

\subsection{Scatter-gather concatenation}\label{sec:scatter_gather}
The block size can have a significant impact on data transfer throughput. Yet, the size of a page, RNTuple's data chunk unit, is fixed when the dataset is generated. The default page size (64\,kB) is reasonable for a local filesystem; however, DAOS may benefit from larger chunks to offset network overhead, especially over RDMA~\cite{5158468}.
To circumvent this limitation, we introduce a scatter-gather-based mechanism to splice many pages within the same page group that are, then, transferred as a single unit.

\section{Evaluation}
In this section, we present the results of the experimental evaluation after applying the improvements described in Section~\ref{sec:proposal}. The experiments were all conducted with the hardware and software environments described below.

\begin{description}
\item[Hardware platform.] Hewlett-Packard Enterprise Delphi cluster, comprising two servers and six client nodes interconnected via InfiniBand fabric. A single client node was used.

\rule{1ex}{0pt}%
\emph{Server nodes.} 4$\times$ Intel Xeon Gold 6240M CPU (18 physical cores) running at 2.60\,GHz, with 24.75\,MB of L3 cache and 185\,GB of DDR4 RAM and hyper-threading SMT enabled. These nodes are equipped with a Mellanox MT28908 ConnectX-6 InfiniBand adapter.

\rule{1ex}{0pt}%
\emph{Client node.} 2$\times$ Intel Xeon E5-2640 v3 CPU (8 physical cores) running at 2.60\,GHz, with 20\,MB of L3 cache and 131\,GB of DDR4 RAM, and hyper-threading enabled. High-speed interconnection is available through a Mellanox MT27800 ConnectX-5 InfiniBand adapter.

\item[Software.] The operating system is Red Hat Enterprise Linux 8.4 (kernel 4.18.0-305), employing daos-2.2.0 (ofi+verbs provider), and libfabric 1.15.1.

\item[Test cases.] To evaluate the proposed method, we used the LHCb Run 1 open data B2HHH (B meson decays to three hadrons) dataset comprising 26 columns, out of which 18 are read for an analysis that culminates in a histogram for the B mass spectrum. The original dataset contains 8.5\,M events and amounts to an uncompressed size of 1.5\,GB. To simulate the use case for HPC clusters, the dataset is artificially concatenated 10-fold to a total of 15 GB.
\end{description}

We measured both \emph{(i)} the transfer rate for importing the entire dataset into DAOS, and \emph{(ii)} the end-to-end analysis throughput, i.e., from storage to histogram, for the following scenarios.
\begin{description}
\item[Baseline:] uses an experimental backend that predates the improvements hereby proposed. It employs the na\"ive object-per-page mapping between RNTuple pages and DAOS objects.

\item[Current (object-per-page):] with the improvements proposed in Section~\ref{sec:proposal} and the previous, object-per-page mapping.

\item[Current (locality-driven):] with the proposed improvements and mapping that exploits co-locality.

\item[Target:] same as above, but the optimization described in Section~\ref{sec:scatter_gather} has been enabled and using a target transfer size of 1\,MiB.
\end{description}

\begin{figure}[ht]
    \centering\vspace{-1em}\resizebox{\textwidth}{!}{\input{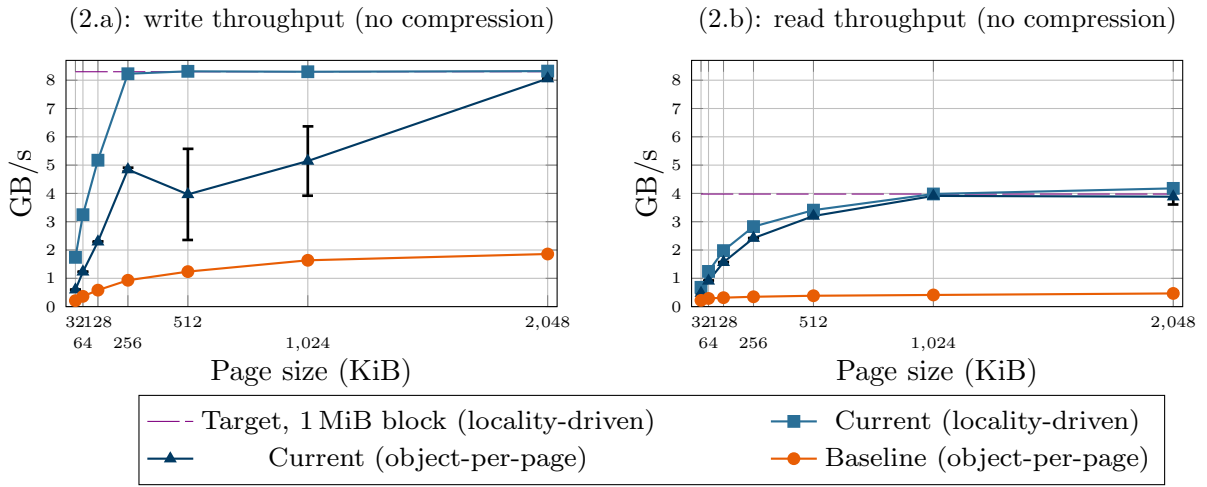}}%
    \vspace{-1em}
    \caption{LHCb analysis throughput across different mappings. Note the wide error bars for the object-per-page mapping. One possible explanation is the impossibility of coalescing requests and the high pressure on the operation queue.}\label{fig:daos_plot_detailed}
\end{figure}

\begin{figure}[ht]
    \centering\resizebox{0.93\textwidth}{!}{\input{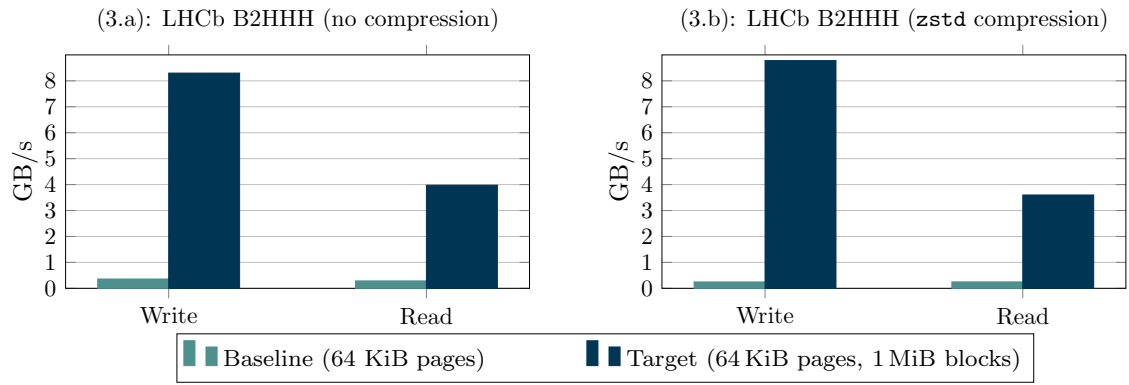}}%
    \vspace{-1em}
    \caption{Throughput comparison for native 64\,KiB pages with the baseline and with the current backend, with 1\,MiB concatenation enabled.}\label{fig:daos_compare}
\end{figure}

Figure~\hyperref[fig:daos_plot_detailed]{\ref*{fig:daos_plot_detailed}.a} and~\hyperref[fig:daos_plot_detailed]{\ref*{fig:daos_plot_detailed}.b} show, respectively, the measured write throughput in GB/s while importing data and the read throughput during the analysis, both averaged over five runs. These plots compare performance w.r.t. various page sizes in each of the scenarios described before. The best results are achieved by using the locality-driven mapping together with page sizes of around 1\,MiB or more. The horizontal dashed line corresponds to splicing pages up to a target size of 1\,MiB.
Figure~\ref{fig:daos_compare} shows a comparison of the read/write throughput attained by both versions of the backend, before and after the improvements, for a dataset stored either uncompressed or using \textit{zstd}. The results demonstrate, respectively for write and read scenarios, $4.4$ to $9\times$ and $3$ to $10\times$ better performance with the proposed method across all page sizes.

\section{Conclusion and future work}
The results shown in the experimental evaluation demonstrate that, with the latest proposed developments, RNTuple's DAOS backend is ready for high-throughput analyses in HPC data centers. In particular, we measured, respectively, a $9$ and $4.3\times$ improvement in the write and read throughputs for the default 64\,KiB page size, which directly translates into much faster data import and end-to-end analysis. By fully leveraging scatter-gather lists, the backend can target transfer sizes associated with higher throughput independently from the page size.

As future work, we plan to extend the evaluation with multi-process and distributed analysis with ROOT's Distributed RDataFrame in order to saturate the capacity of the link layer. Moreover, significant development efforts are underway to natively support the S3 object store.

\section*{Acknowledgments}
This work benefited from support by the CERN Strategic R\&D Programme on Technologies for Future Experiments CERN-OPEN-2018-006 and the Intel--CERN openlab collaboration.
Access to the hardware for the experimental evaluation was provided by Hewlett-Packard Enterprise.

\printbibliography
\end{document}